# RECTANGULAR WAVEGUIDE HOM COUPLER FOR A TESLA SUPERSTRUCTURE

A.Blednykh, D.Kostin, V.Kaljuzhny, M.Lalayan, O.Milovanov, A.Ponomarenko, N.Sobenin, A.Sulimov, D.Trubin, A.Zavadzev, MEPhI, Moscow, Russia
M.Dohlus, DESY, Humburg, Germany

*Abstract*

Some modifications of a Rectangular Waveguide HOM couplers for TESLA superstructure have been investigated. These RWG HOM couplers are to be installed between the cavities of the superstructure and also at the both ends of it. We investigated a RWG HOM coupler attached to the beam pipe through the slots orientated along beam pipe axis (longitudinal slots), perpendicular to it (azimutal slots) and at some angle to this axis. For dipole modes of both polarizations damping two RWG in every design were used. This paper presents the results obtained for scaled-up setup at 3 GHz at room temperature. The advantages of HOM coupler with longitudinal slots for damping dipole modes and compact HOM coupler with slots at some angle to the axis are shown. Arrangement of HOM coupler in cryostat and heating due to HOM and FM losses are presented. Calculations and design of the feeding RWG coupler for superstructure are also presented.

## 1 INTRODUCTION

We have chosen for mock-up of superstructure the operational frequency 2.981 GHz instead of 1.3 GHz in order to reduce its dimensions. This led to scaling factor $K_{scl}$=2.293. Alsothe dimensions of all cells except end ones were decreased in 2.293 times as compared with TESLA dimensions [1]. End cells iris diameters were 49.70 mm (114mm at 1.3GHz) and 34.04 mm (78mm at 1.3GHz).

We have examined in [2] a variant of set-up consisting of Fundamental Mode (FM) coupler between subcavities and HOM coupler of three RWGs, coupled through azimuthal slots with beam pipe. Such HOM coupler has appeared to be inefficient for dipole modes damping. Instead of it another variant was proposed having FM coupler attached to a beginning of superstructure and HOM device of two RWGs with slots orientation along the beam pipe axis. This HOM coupler is preferable for dipole modes damping but its disadvantage is due to the insufficient place for an RWG placement between two subcavities, and it can be mounted only at the ends of superstructure. The device with the coupling slots revolved to some angle from beam pipe axis is the more suitable for HOM damping device between the subcavities. The rejecter filter was used for the operational mode penetration preventing through this device.

## 2 THE TEST MODULE

We have considered RWG FM coupler design in the beginning of two subcavities coupled through the beam pipe having the diameter 49.70 mm (see Fig.1). FM coupler consisted of 72×13.08 mm$^2$ RWG short-circuited at the one end and coupled to the cavity by the beam pipes with diameter 34.02 mm. A short-circuits consist of cylindrical surfaces with diameter 42.74 mm and two plane surfaces tangential to the latter (see Fig.2). During the calculations done to obtain the desired value of $Q_{ext}$, we changed an angle between two plane surfaces and position of FM coupler $Z_{LC}$. The results of these calculations are presented at Fig.3.

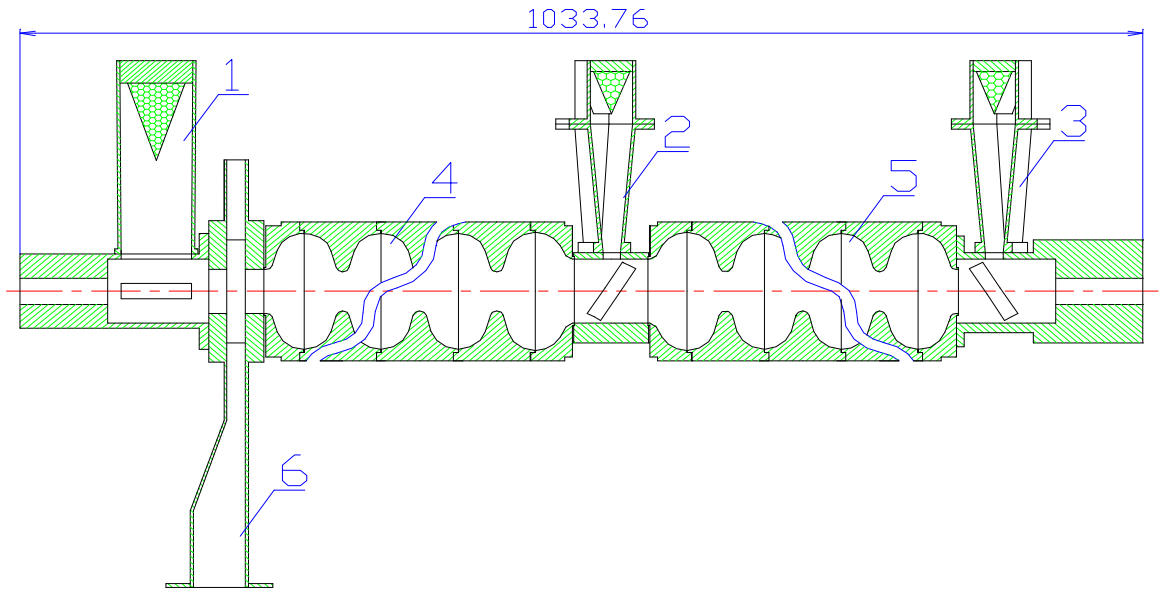

Figure 1: TESLA S-band test module.1,2,3-HOM couplers; 4,5-cavities; 6-FM coupler.

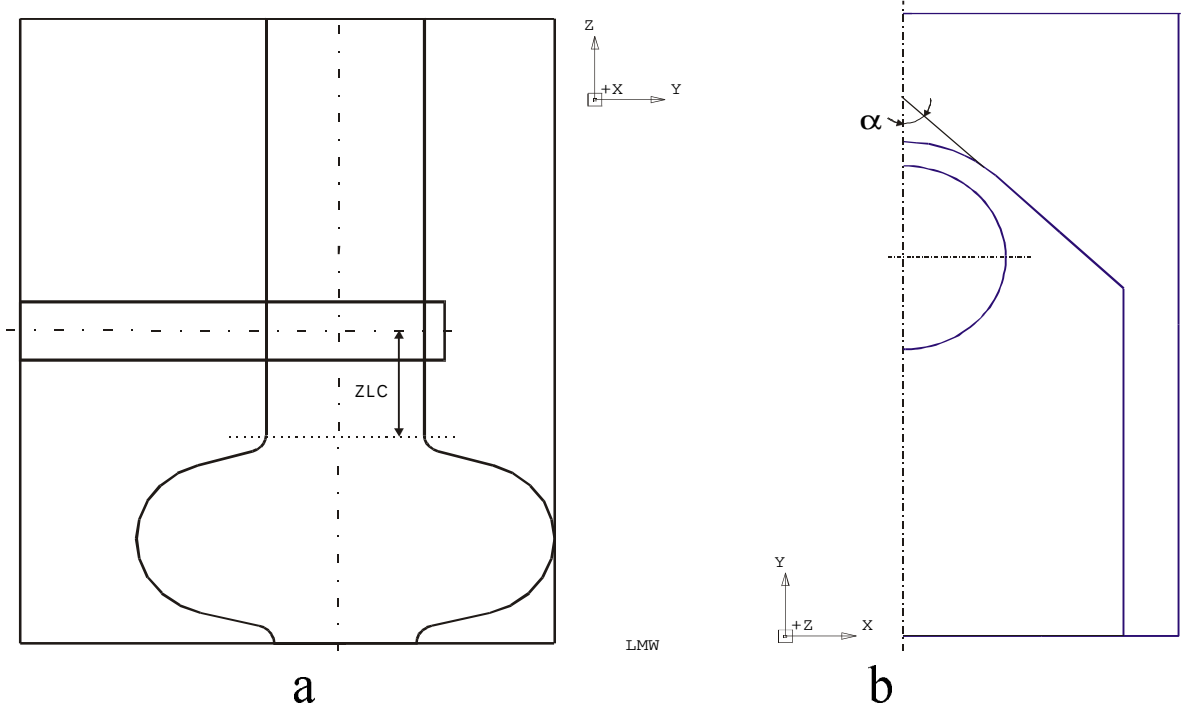

a b

Figure 2: FM coupler with HOM damper

It is known that frequency scaling does not affect on external quality factor $Q_{ext}$ and in case of 2×7-cells cavity (and also 4x7 cells) it must be equal $Q_{ext}$ =3.36×10$^6$ [3]. In the case of used for calculations model concisted of one cell coupled with the RWG coupler $Q_{ext}$ must be 240000.

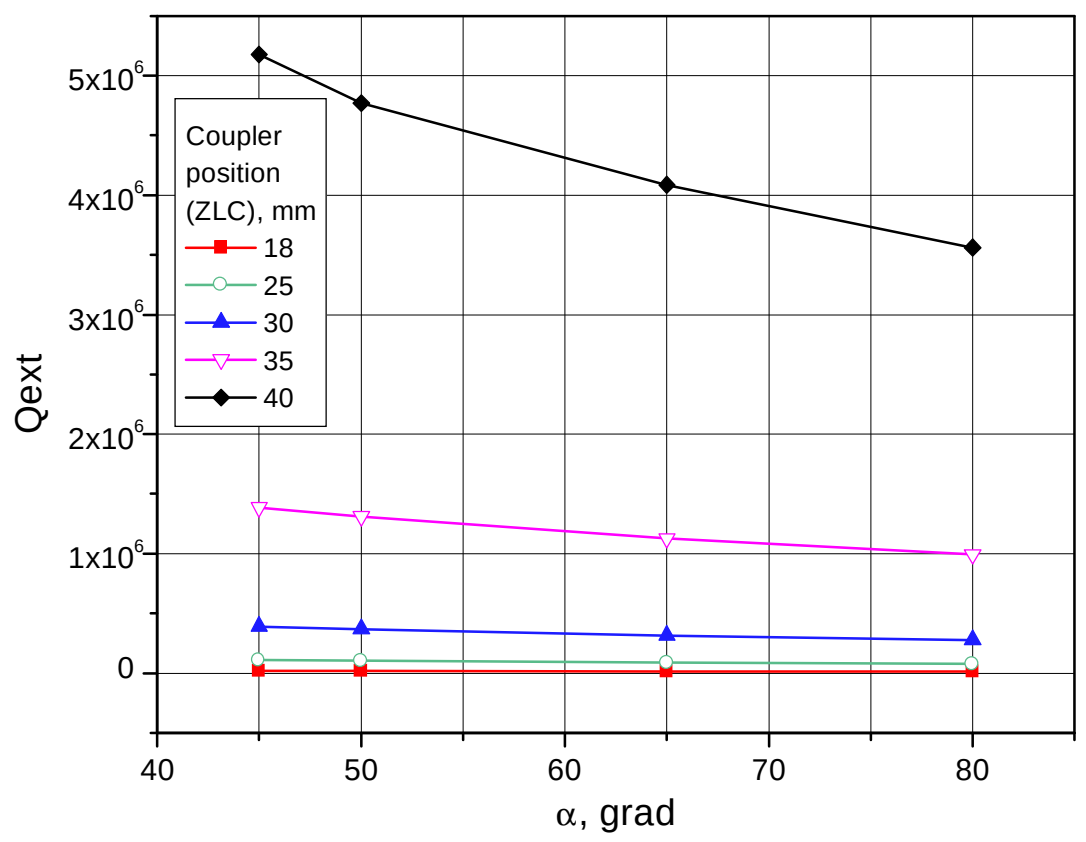

Figure 3: $Q_{ext}$ vs. $Z_{LC}$ and α

For the HOM damping we have considered two RWGs of 46x12 $mm^2$ crossection coupled with the beam pipe diameter 49.70 mm through the slots along beam pipe axis (see 1 at Fig.1). Dipole HOMs excited in the cavity creates $H_{11}$-like fields in the beam pipes connecting subcavities. For this beam pipe mode the ratio $H_{zmax}$ and $H_{\varphi max}$ components is more then 1 in wide frequency range and wide sides of the rectangular waveguides must be orientated along the beam pipe. RWGs in this coupler have a strong coupling with dipole modes, provide strong HOM damping with no coupling with the fundamental mode and other monopole modes.

But for this construction with longitudinal slots it is difficult to install the RWG between two cavities, because the beam pipe is shorter than wide RWG wall. Therefore we had to use waveguide transition or the construction of two rectangular waveguides with dimensions 46x10 $mm^2$ and the slots rotated by some angle in respect to the beam pipe axis (see 2 at Fig.1).

## 3 THE HIGH ORDER MODES INVESTIGATIONS

In order to restrict the multi-bunch phenomena due to wakefields, the higher order modes of the TESLA cavity must be damped down to the certain $Q_{ext}$ level of $10^4$ - $10^5$. These $Q_{ext}$ measured in S-frequency band must be the same as $Q_{ext}$ in TESLA frequency band.

We have determined the $Q_{ext}$ for different RWG HOM coupler using the results of quality factor measurements for the two cavities with beam pipe and without HOM coupler $Q_1$, and with HOM coupler $Q_2$. The external Q-factor may be calculated now from

$$Q_{ext}=Q_1Q_2/(Q_1 - Q_2)$$

The results of measurements using of bead pulling technique for all three HOM couplers are summarized in Table 1 and Table 2. The asterisk * marks those HOMs, for which $Q_1$=$Q_2$. Dash designates absence of excitation at the appropriate HOM coupler presence.

## 4 RWG HOM COUPLERS IN TESLA CRYOSTAT

We considered the possibility to arrange the RWG HOM coupler in TESLA cryostat [4]. Three RWG HOM couplers are shown in Fig.4. This coupler consists of two 110×20$mm^2$ RWGs for dipole HOMs and 70×20$mm^2$ RWG for monopole ones. To decrease longitudinal dimension we use tapered waveguide transitions from 110mm to 95mm and from 20mm to 10mm. All transmissions have 70mm length. The longitudinal position of 110x20$mm^2$ RWG was chosen due to following reason. The lowest dipole HOM excites $H_{11}$-like electromagnetic field in the beam pipe tube. At the frequencies of these HOMs the ratio Hz/Hφ = 2.5 - 6.5. Thus Hz component of the magnetic field provides strong coupling of lowest dipole HOMs with $H_{10}$ mode in RWG

Fig.4 shows RWGs location in the cryostat. The initial part of each RWG of 90 mm length (from beam pipe tube to flange) is made of Nb and is superconducting (2K). The final part of each RWG is made of stainless steel covered by copper. The wall thickness of latter part equals 0.2 mm and has the length of about 1 m. These waveguides are terminated by the matched loads, which have 70K temperature.

Another variant has two RWG HOM couplers have its RWGs rotated through 45-60$^0$. It reduces the longitudinal RWG size and permits coupling with both monopole and dipole HOMs including fundamental mode. To reduce the coupling with fundamental mode, we used narrow frequency band rejection filter tuned on resonance with fundamental mode fields.

The calculation results of heat loading for these two variants are present in Table 3.

## 5 CONCLUSION

The RWG HOM coupler with longitudinal slots provide rather strong connection with dipole modes of the subcavity. Besides this coupler does not need the rejection filter, which is required by the coaxial HOM coupler to avoid the damping of the fundamental mode.

Table 1. Comparison RWG HOM couplers

| HOM | Frequency, MHz | | $Q_0$ $x10^3$ | $Q_{ext}, 10^3$ HOM coupler type. | | |
|---|---|---|---|---|---|---|
| | Model | TESLA | | ⊥ axb=46x12mm | II axb=46x12 mm see Fig.5 | // axb=46x10mm see Fig.2 |
| $H_{111}$ | 3744.47 | 1632.90 | 9.6 | 451 | 17.5 | 48 |
| | 3784.07 | 1650.20 | 8.1 | * | 2.0 | 18 |
| | 3840.13 | 1674.65 | 12.8 | 260 | - | 13.6 |
| | 3843.10 | 1675.94 | 7.0 | * | 1.75 | 9.9 |
| | 3918.58 | 1708.89 | | * | - | - |
| | 3922.84 | 1710.75 | | 36.0 | 5.44 | 5.44 |
| | 4000.60 | 1744.62 | 10.2 | * | - | 0.74 |
| | 4009.21 | 1748.37 | 11.5 | 209 | 32.6 | 7.9 |
| | 4074.72 | 1776.94 | 10.1 | * | - | 0.6 |
| | 4089.85 | 1783.54 | 11.6 | 180 | 13.8 | 5.0 |
| $E_{110}$ | 4127.07 | 1799.77 | 10.2 | 105 | - | 3.1 |
| | 4131.77 | 1801.92 | 9.9 | * | 17.7 | 15.9 |
| | 4229.48 | 1843.60 | 16.9 | * | - | 1.3 |
| | 4233.52 | 1845.99 | 16.3 | * | 7.6 | 19.6 |
| | 4269.37 | 1861.59 | 15.7 | * | - | 1.1 |
| | 4273.83 | 1863.61 | 12.2 | * | 20.1 | 37.4 |
| | 4298.55 | 1874.40 | 16.7 | * | - | 15.3 |
| | 4302.03 | 1875.92 | 16.8 | * | 38.6 | 124 |
| | 4317.11 | 1882.60 | 14.5 | * | 2.1 | 2.9 |
| | 4319.46 | 1883.60 | 15.7 | * | 2.1 | 29.9 |
| | 4327.33 | 1887.32 | 16.7 | * | 3.1 | 4.1 |
| | 4330.72 | 1888.58 | 14.7 | * | - | * |
| | 4332.00 | 1889.18 | 14.5 | 90.6 | 4.9 | 10.8 |

Table 2. Comparison RWG HOM couplers for $E_{011}$ mode

| F, MHz | $Q_0$ $x10^3$ | \|\| axb= 46x10 mm see Fig. 5 | | ⊥ axb= 28x12.5 mm | | \|\|+⊥+\|\| \|\|-axb= 46x12 mm ⊥-axb= 28x12.5 mm | |
|---|---|---|---|---|---|---|---|
| | | $Q_L$ $x10^3$ | $Q_{ext}$ $x10^3$ | $Q_L$ $x10^3$ | $Q_{ext}$ $x10^3$ | $Q_L$ $x10^3$ | $Q_{ext}$ $x10^3$ |
| 5481.83 | 15.4 | 14.5 | 248.1 | 12.7 | 72.44 | 14.7 | 323.4 |
| 5502.40 | 5.1 | 4.7 | 59.9 | 4.75 | 69.2 | 4.7 | 59.92 |
| 5509.82 | 11.9 | 8.9 | 35.3 | 5.55 | 10.4 | 8.7 | 32.35 |
| 5539.50 | 9.0 | 5.9 | 17.13 | 3.85 | 6.73 | 4.73 | 9.97 |
| 5577.20 | 9.7 | 4.6 | 8.75 | 3.35 | 5.12 | 2.53 | 3.42 |
| 5599.93 | 12.4 | 5.0 | 8.38 | - | - | - | - |
| 5618.70 | 11.2 | 3.4 | 4.88 | 2.5 | 3.22 | 1.73 | 2.05 |
| 5633.00 | 14.0 | - | - | - | - | - | - |
| 5643.90 | 14.5 | 4.1 | 5.72 | 4.0 | 5.52 | 1.86 | 2.13 |

Table 3. Heat loading of 2K region per 4×7 cells superstructure

| Version | RWG dimensions, axbxt, $mm^2$ | Nonsuperconducting RWG length, mm | Static losses per one RWG 70→2K, mW | Fundamental mode losses per one RWG, mW | Number of RWGs | Losses per superstructure, mW |
|---|---|---|---|---|---|---|
| 1 | 110x20x0.2 | 1000 | 13 | 0 | 2x4 | 148 |
| | 70x20x0.2 | 1000 | 9 | 2 | 1x4 | |
| 2 | 110x20x0.2 | 1000 | 13 | 4 | 2x5 | 170 |

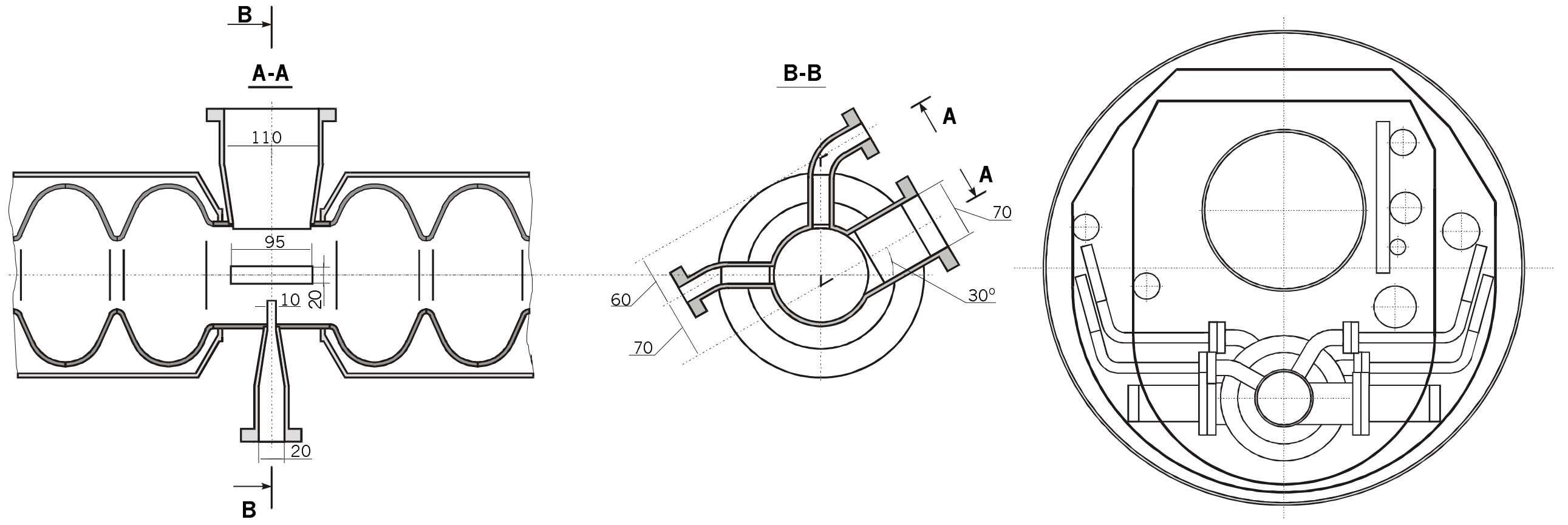

a b

Figure 4. Three RWG HOM coupler locations in the cryostat.